\documentstyle[12pt,aasms4]{article}
\received{1998 October 28}
\accepted{1999 February 2}

\begin{document}
\title{THE STELLAR POPULATION OF THE M31 SPIRAL ARM AROUND OB ASSOCIATION A24}
\author{Keiichi Kodaira\altaffilmark{1}, Vladas Vansevi\v{c}ius\altaffilmark{1,2},
Motohide Tamura\altaffilmark{1}, and Satoshi Miyazaki\altaffilmark{1}}

\altaffiltext{1}{National Astronomical Observatory, 2-21-1 Osawa, Mitaka,
Tokyo 181-8588, Japan;\\
kodaira@cc.nao.ac.jp, vladasvn@cc.nao.ac.jp, tamuramt@cc.nao.ac.jp,\\
satoshi@merope.mtk.nao.ac.jp}

\altaffiltext{2}{Institute of Physics, Go\v{s}tauto 12, Vilnius 2600, Lithuania;
wladas@itpa.lt}

\begin{abstract}
A study of the stellar population of the M31 spiral arm
around OB association A24 was carried out based on the
photometric data obtained from deep $V$ and $JHK$
imaging. The luminosity function was obtained for $-7 \lesssim
M_{bol} \lesssim -3.5$ by applying the extinction correction
corresponding to $A_V=1$ and the bolometric correction
$BC_K$ as an empirical function of $(J-K)_0$. In
comparing the observed color-luminosity diagrams with
semitheoretical isochrones modified for the dust-shell
effects, we found the young population of $t \lesssim 30$ Myr with
supergiants of
$M_{bol} \lesssim -5$, the bulk of the intermediate-age population of
$t \sim0.2 - 2.5$ Gyr with bright asymptotic giant branch (AGB) stars of
$-5 \lesssim M_{bol} \lesssim -4$, and old populations of $t \gtrsim 3$
Gyr with
AGB and red giant branch (RGB) stars of $M_{bol} \gtrsim -4$. The average star
formation rate was estimated to be $\sim 1.8 \times 10^4$
$m_\odot$ Myr$^{-1}$ and $\sim 0.7 \times 10^4$ $m_\odot$ Myr$^{-1}$
per deprojected disk area of 1 kpc$^2$ from the number
density of B0 V stars around $M_V=-4.0$ (age $\sim 10$
Myr) and the number density of bright AGB stars around
$M_{bol} = -4.3$ (age $\sim 1$ Gyr), respectively. A
study of the local variation in the $V$ and the $J$ and
$H$ luminosity functions revealed a kind of
anticorrelation between the population of the young
component and that of the intermediate-age component
when subdomains of $\sim 100$ pc scales were concerned.
This finding suggests that the disk domain around the
A24 area experienced a series of star formation episodes
alternatively among different subdomains with a timescale
of a few spiral passage periods. Brief
discussions are given about the interstellar extinction
and about the lifetimes of bright AGB stars and the
highly red objects (HROs) in the same area.
\end{abstract}


\section{Introduction}

The Andromeda galaxy (M31) is the nearest galaxy which has
a morphological
type close to that of the Milky Way, and is most suitable to provide
supplementary data for understanding the detailed
galactic structures such as spiral arms and related
star forming regions. With this motivation in mind, the
present authors carried out deep near-infrared stellar
photometry around the OB association A24 $({\rm
R.A.}=0^h39^m18^s$, ${\rm Decl.}=+40^{\circ}41'
~\lbrack 1950.0 \rbrack)$ near the 7 kpc spiral arm of
M31 (see Fig. 1), and published the basic photometric
data separately (Kodaira et al. 1998a, hereafter Paper I). In the
present paper we carry out a population study of the
stars around A24 based upon the near-infrared data in
Paper I and the new $V$-band data.

An automated search for OB associations in M31 was
carried out by Magnier et al. (1993) based upon the CCD
survey data of Magnier et al. (1992), and Haiman et al.
(1994) analyzed the properties of a dozen OB
associations selected from the list of Magnier et al.
(1993). In order to improve the data quality, Magnier et
al. (1997) undertook $UBV$ photometry using the Hubble Space Telescope (HST)
WFPC for 15 fields related to the OB associations A41,
A42, and A48. They found that the median ages of the
bright blue stellar components were $\log t$ (years)$=7.0 - 7.5$
with an average $\log t \sim 7.3$, and that the
total interstellar extinction of individual stars was
highly variable over a range $A_V=0 - 3$ with group
median values of $A_V \simeq 0.6 - 1.3$ and an average
of $A_V \sim0.9$ assuming $R_V=3.1$.

The present target, OB association A24 (area $\sim
2'\times1.'5$), is composed of a scattered group of
moderately faint stars, with a core size of about 0.1
kpc (van den Bergh 1964; see Fig. 1). This core includes
open clusters C202 and C203, both small with bright
stars, near the rim of the dark cloud D221 (area
$\sim0.'8\times0.'8$), which appears at the central part
of A24 (Hodge 1981). Efremov, Ivanov, \& Nikolov (1987)
identified the original A24 as A24-1, and additionally
defined new associations A24-2 and A24-3 on the
southeast outside A24-1. They noticed that A24-1 was
a stellar complex which contained at least five internal
groupings. Other dark clouds, D205, D215, and D253, and
a faint open cluster C204 are seen in the vicinity of
A24.

The clouds D221 and D253 compose a part of the spiral
dark lane which appears to branch off inward from the
10 kpc main dark lane on the south side of M31. The HI
distribution maps by Unwin (1980) and Brinks \& Shane
(1984) show that an HI spiral arm runs through D221, but
that the association A24 including the cloud D221 is
located at a local minimum of HI distribution along the
spiral arm. The $IRAS$ maps for the emission at 60
and 100 $\mu m$ by Xu \& Helou (1996a) also show that the
spiral arm is weakening in the vicinity of A24. Clear
signs of diffuse H$\alpha$ emission were seen in the A24
area in the H$\alpha$ survey maps of Pellet et al.
(1978) and of Devereux et al. (1994). The H$\alpha$
features were more clearly identified by Magnier et al.
(1995) in their search for supernova remnants (SNRs) in M31.
Their study revealed that a superbubble of $38''$
($\sim 130$ pc) diameter developed around the open
clusters C202 and C203, with two possible SNRs in the
neighborhood.

These features of A24 (with a superbubble and at a local
minimum of the HI spiral arm) highly resemble to those of
A41a, which was investigated by Magnier et al. (1997).
They found A41a to have a median age of $\log t \sim7.3$
and a median extinction of $A_V \sim0.6$. The foreground
reddening in the Milky Way toward M31 was estimated at
$E_f(B-V)=0.16$ (Sandage \& Tammann 1968),
$E_f(B-V)=0.07$ (Humphreys 1979), or $E_f(B-V)=0.08$
(Burstein \& Heiles 1984). Hodge \& Lee (1988) found
internal reddening $E_i(B-V)=0.21$ for A24. Our
reanalyses of the blue stars in the Hodge \& Lee (1988)
field using their $UBV$ data and the $BVRI$ data of
Magnier et al. (1992) resulted in $E_i(B-V)\simeq 0 - 1$
for individual stars, with a median value of
$E_i(B-V)=0.23$. The extinction seems to be highly
inhomogeneous over the A24 field, as was found by Magnier
et al. (1997) for other associations.

Since we are mainly dealing with the near-infrared data,
the uncertainty and the inhomogeneity of the extinction
little affect the interpretation about the global
characteristics of the stellar population. Accordingly,
in the present population study we assume an average
extinction of $A_V=1.0$, which corresponds to
$A_K=0.11$, $E(J-K)=0.17$, $E(H-K)=0.07$, and
$E(B-V)=0.32$ when the extinction law for $R_V=3.1$ is
adopted (Mathis \& Cardelli 1990). The main conclusions
from the present study are hardly affected by an
uncertainty of $\Delta A_V=\pm0.2$. Since the
photometric zero point of the $K$ magnitude was adjusted
in Paper I by fitting the observed main sequence to the
unreddened one in the $(J-H)$ versus $(H-K)$ diagram, we
need in the case of $A_V=1.0$ to subtract corrections
$\Delta(J-H)=0.10$, $\Delta(J-K)=0.10$, and $\Delta
K=0.07$ from the photometric data given in the Paper I,
for the comparison with (unreddened) theoretical values.
The colors with subscript zero and the absolute
magnitudes are for the dereddened values, while the
original data in the Paper I photometric system appear
without a subscript zero in the present paper.

In the following, a description of the observational
material is given in \S\ 2, and a population study is
presented in \S\ 3. We will investigate the local
variations of stellar population and of interstellar
extinction in \S\ 4. Discussions of star
formation and cloud structure in the A24 area are
given in \S\ 5. Brief discussions are also given in \S\
5 of the lifetimes of bright AGB stars and of the
HROs identified by Kodaira et al. (1998b).

\section{Observational Material}

Our near-infrared data consist of two sets, which were
both obtained with the QUIRC imager attached to the
University of Hawaii $2.2$ m telescope at Mauna Kea. One set is the $JHK$ data
obtained in 1995 September with the f/31 tip-tilt
secondary mirror, and their point-spread function (PSF) is as good as
FWHM $\simeq 0.''4$. This set of high-quality data covers the
central field of $2'\times2'$ around D221 and reaches
down to $J=21.3$, $H=20.5$, and $K=19.6$. The other set
is the $JH$ data, which were obtained in 1996 December
with the f/10 normal secondary mirror and covered a
$3'\times 3'$ field around the f/31 field. The set of the
f/10 $JH$ data has a PSF of FWHM $\simeq 0.''9$ and
reaches down to $J=20.0$ and $H=19.0$.
More details about the near-infrared photometry are to
be found in Paper I.

The $V$-band image was obtained on 1995 September 23
using the University of Hawaii 8K mosaicked CCD camera at the f/3.8 primary
focus of the Canada-France-Hawaii Telescope (CFHT).
The mosaic of eight $2{\rm K} \times 4{\rm K}$
front-illuminated detectors with 15 $\mu m$ pixels ($0.''23$ pixel$^{-1}$) covered
a $0.^{\circ}52 \times 0.^{\circ}52$ field of the south
part of M31 including the spiral arms around A24. The
900 s exposure frame was debiased using a 900 s dark
frame and flattened using the sky flat field, which was
produced by taking a median of 20 frames. The final
object frame has stellar images of FWHM $\sim0.''9$. A
$7'\times7'$ field around A24 was carefully measured to
yield a $V$-band magnitude for about 29,000 stars down to
$V\sim24$. The photometric calibration was made by
referring to the M31 survey data by Magnier et al.
(1992). The resulting $V$-band luminosity function is
given in Figure 2, which shows 3 components
corresponding to supergiants ($M_V \lesssim -6$), OB
main-sequence stars ($-6 \lesssim M_V \lesssim -3.5$), and RGB and AGB
stars $(-3.5 \lesssim M_V)$.

We will use the high-quality f/31 $JHK$ data for the
population study for the $2'\times2'$ central field, and
the f/10 $JH$ data and the f/3.8 $V$ data for the study
of local variation in the stellar population and the
interstellar extinction over the $3'\times3'$ and the
$7'\times7'$ fields, respectively.

\section{Population Study}

We begin with the population study of the 3139 stars
observed in the $JHK$ bands in the $2'\times 2'$ central
field of A24 (see Fig. 1b), whose effective area is
$\sim 3$ arcmin$^2$ because of mosaicking seams.
The $K$-band luminosity, $M_K$, was transferred to the
bolometric luminosity, $M_{bol}$, using the empirical
relation between $(J-K)_0$ and the $K$-band bolometric
correction $BC_K$, which were derived by us based upon
the data in Whitelock et al. (1995), Ku\v{c}inskas
(1998), and Montegriffo et al. (1998). The details of
the transformation are given in Appendix A.

The conventional value of the distance modulus $m-M=24.2$
(Welch et al. 1986; Pritchet \& van den Bergh 1987;
Huterer, Sasselov, \& Schechter 1995) is adopted for M31
throughout this paper, although Stanek \& Garnavich
(1998) and Holland (1998) are proposing a larger value,
$m-M=24.47$, based upon HST $VI$-band data about the
"red clump" stars and the RGB stars, respectively.
Our main conclusions in the following are hardly
affected by the ambiguity of $\Delta(m-M)= +0.3$. The
resulting luminosity function is shown in Figure 3. The
luminosity function shows three components that
correspond to the few brightest objects with $M_{bol} \lesssim -6.5$,
bright stars with $-6.5 \lesssim M_{bol} \lesssim -5$, and the dominant
component with $M_{bol}\gtrsim-5$. The brightest part of this
luminosity function for $M_{bol} \lesssim -6$ is subject to large
statistical uncertainty because of the scarceness of
these stars in the $2'\times2'$ field. The change of
slope that is seen in the $K$ magnitude distribution
around $K\sim 17.5$ (see Fig. 4b in Paper I) is not clear in Figure 3.

In Figures 4 and 5 we compare the observed two-color
diagram (TCD) and color-magnitude diagrams (CMDs) with
the isochrones, of which the original theoretical ones
are based on those of the Padova group (Bertelli et al.
1994, hereafter Padova 94). We have plotted only 844
stars of $K<18.0$ in Figure 4 because the colors of the
fainter stars are increasingly subject to the
photometric errors (see Paper I). In the TCD the Padova
94 isochrones for various ages and metallicities are
almost degenerated, and cannot cover the upper domain
where the observed red stars of $J-H>0.7$ are
distributed, while the tracks for RGB and AGB stars
along $J-K\simeq1.2$ extend far beyond the populated
domain. This discrepancy between the observed stellar
distribution and the  Padova 94 isochrones in TCD may
mainly be attributed to the following two causes. One
apparent cause is connected to the difficulties in
transforming the theoretical HR diagram ($\log L$, $\log T_{eff}$)
into the color-magnitude diagram. Because of the
lack of a sufficient spectral library for evolved cool
stars and the uncertainty in the correspondence between
the spectral types and the theoretical parameters ($g$,
$T_{eff}$) for these stars, Padova 94 had to invoke
various simplifications in producing spectral energy
distributions (SEDs) for stars of low effective
temperature ($T_{eff}<3500$ K) and low surface gravity
($\log g<3$). The second cause is neglect of the effects
of the dust shells which are probably ejected during the
AGB evolution, which was not taken into account in the
Padova 94 models. Bressan, Granato, \& Silva (1998)
modeled the effects of the dust shells and
reproduced fairly well the observed color characteristics of the
$IRAS$ AGB stars in the South Galactic Cap (Whitelock et
al. 1995). In doing so, they adopted the underlying
photospheric radiation roughly corresponding to those
halfway on the Padova 94 evolutionary tracks of AGB
stars with colors $(J-H)_0\simeq 0.9 - 0.7$ and
$(H-K)_0\simeq 0.2 - 0.5$, and modified them for the
dust-shell effects. If we apply the dust-shell
corrections to the photospheric radiation corresponding
to the latest stage of the Padova 94 AGB evolution,
which has almost the constant colors of
$(J-H)_0\simeq0.5$ and $(H-K)_0\simeq0.8$, the resulting
colors do not match the observed ones; they are too
red in particular in $(H-K)_0$. The deviation of the
original Padova 94 isochrones from the observed stellar
distribution in CMDs (see Fig. 5) also indicates the
necessity of modifying the colors of the evolved cool
stars. In extrapolating SEDs along with varying colors
$(V-K)$ and $(J-K)$ in Padova 94, there might have been
some anomalous behaviors in the $H$ band which led to
the unexpected extent of the tracks for AGB and RGB
stars in TCD, namely, the $H$-band flux might have
become too small.

The supplement to the SED library and/or the improvement
of the correspondence between the spectral types and the
physical parameters are outside the scope of the present
observational work. Here we adopt the practical way to
empirically $terminate$ the color evolution of AGB stars
halfway along the Padova 94 tracks and to modify the
colors by correcting them for the dust-shell effects.
For the present purpose we adopted $(J-H)_0=0.7$ and
$(H-K)_0=0.3$ as the average $terminal$ colors for the
AGB evolution (but not for the RGB evolution) and
applied the dust-shell corrections to them. The
dust-shell corrections normally become effective for
high-luminosity stars of $M_{bol} \lesssim -3.8$. The details of
the dust-shell models are described in Appendix B. Note
that the present dust-shell model assumes the silicate
grains, which have different extinction properties from
those for the standard Galactic interstellar absorption,
and that the combined effects of the silicate absorption
and the thermal emission from the silicate dust-shell
together mimic the standard Galactic reddening in Figure
4. The main isochrones in Figures 4 and 5 are the
semitheoretical ones which were derived this way, and
the original Padova 94 isochrones are shown in the
insets for comparison. We have adopted the isochrones
for the helium and metal fractions $Y=0.25$ and
$Z=0.008$ for the ages of 5 and 10 Gyr, and $Y=0.28$
and $Z=0.02$ for younger ages. The modified isochrones
are still partially degenerated in TCD, and the tracks
for the RGB stars remain the same, for which no
modifications have been applied. These semitheoretical
isochrones in CMDs are useful to lend us a guiding
reference, while the present population study of AGB
stars relies on $M_{bol}$ rather than on the detailed
colors in their late evolutionary phase.

By investigating the CMDs, we find that the youngest
population has an age of $t \lesssim 30$ Myr when a
metallicity similar to that of the solar value,
$Z\sim 0.02$, is adopted. If $Z\sim 0.05$ is adopted, the
age becomes slightly younger. The brightest stars in the
open clusters C202 and C203 (the two most luminous) must
be as massive as $m \gtrsim 60$ $m_\odot$ and as young as
$t \lesssim 4$ Myr if measured images are actually
resolved single stars. Note that the youngest main-sequence
stars of $M_{bol}\gtrsim-5.5$ are below the detection
limit in the $K$ band. Judging from the broad ridge
running through the bright AGB stars of $M_{bol}\sim-5$
down to $M_{bol}\sim-3.5$ in CMDs, we find the ages of
the bulk of the bright AGB population ($-5 \lesssim M_{bol} \lesssim
-4$) to be
$t\sim 0.2 - 2.5$ Gyr for $Z=0.02$, as normally expected
for stars of about $\sim1.5 - 4$ $m_\odot$. The
component of $M_{bol}\gtrsim-4$ includes the old population of
$t\sim 3 - 10$ Gyr having $Z<0.02$. We suspect that the
core part of the lower right extent of the stellar
distribution in CMDs (Fig. 5) may include RGB and AGB
stars of $M_{bol}\gtrsim-3.0$ which are suffering from the
local absorption of $A_V\sim 1 - 3$ in M31 in addition to
the average $A_V=1$ adopted above. The reddening effects
for these faint stars, however, are not distinct in TCD
because the size of the observational errors rapidly
increases with decreasing stellar brightness toward the
detection limits (see Paper I). For reference we have
indicated in Figure 5a a $dereddening$ vector for
$A_V=5$ according to the Galactic reddening law.

The observed bright AGB stars of $M_{bol} \lesssim -4$ are found
to be distributed in the domain of the Mira variables in
TCD (Bessell \& Brett 1988), corresponding to the redder
group of the South Galactic Cap $IRAS$ AGB stars
(Whitelock et al. 1995). The majority of the observed
objects of $-4<M_{bol}<-3$ shows concentration in the
domain of the semiregular variables in TCD (Kerschbaum,
Lazaro, \& Habison 1996).

The possible contamination by the foreground Galactic stars is expected
to be about 3 in a $1'\times1'$ observed field for $K\leq19$ (see Rich,
Mould, \& Graham 1993) and can be neglected in the present statistical analysis.
On the contrary, judging from Kent's (1989) model and near-infrared
surface photometry by Hiromoto et al. (1983), Battaner et al. (1986),
and Martinez Roger, Phillips, \& Sanchez Magro (1986), the contamination
by the stars in the outer bulge of M31 may not be negligible because of
the inclination angle of M31 at $i=12.^{\circ}5$. The bulge stars,
however, are most probably old enough to be $t>3$ Gyr, and the present
discussions about A24 focused on the stars of $M_{bol} \lesssim -4$ are
hardly affected by their contamination.

\section{Local Variation of Stellar Population}

In order to study possible local variation of the
stellar population around A24, we have derived the
magnitude distributions for individual $20''\times20''$
mesh fields in the $7'\times7'$ $V$-band field as well
as in the $3'\times3'$ $J$- and $H$-band fields (Figs. 6
and 9). Using the number density of stars for $22 \leq V
\leq22.5$ as a criterion, we rather arbitrarily identify
33 subdomains in the $7'\times7'$ field by combining mesh
fields of similar stellar density.  Each of such subdomains has an area
of about 0.8 arcmin$^2$, and they are numbered from 1 to 33, see Figure 6.
There remain mesh fields that are not combined into
subdomains because of relatively high local fluctuation
of stellar density. By analyzing properties of the
$V$-band luminosity functions of these subdomains, we
find that the main features of the luminosity function
can be reproduced by synthesizing a young bright
component, $\psi_1(V)$, and an older faint component,
$\psi_2(V)$, with a varying ratio of $r_V=\psi_1/\psi_2$,
say at $V=22$. The supergiant component seen in Figure 2
$(M_V \lesssim -6.0)$ is too scarce in each subdomain and is
ignored in defining $\psi_1$. The value of $r_V$ ranges
from 0.1 to 1.1, and some of the typical cases are shown
in Figure 7. Both $\psi_1(V)$ and $\psi_2(V)$ can be
approximated in Figures 2 and 7 by a power law, $\log
\psi_i=A_iV+B_i$, with $A_1$ and $A_2$ being $\sim 0.50$
and $\sim 2.24$, respectively. The level constant $B_2$
for the older faint component $\psi_2(V)$ seems to
decrease systematically outward reflecting the density
gradient in the M31 disk: $B_2(No. 14)\simeq
B_2(No. 32)\simeq B_2(No. 1)\times0.7$, and $B_2(No. 22)
\simeq B_2(No. 30)\times 0.95$.

In addition to this main property, many subdomains
reveal a sign of interstellar extinction, that is, a
horizontal shift of the luminosity function toward the
fainter side; $\psi_2(V)=\psi_{2,0}(V-\Delta V)$ with
the subscript zero being for the least absorbed,
reference subdomain of similar population
characteristics. The value of $\Delta V$ ranges from 0
to 0.5, and the typical cases are shown in Figure 8. We
classify the local stellar population by the apparent
extinction $\Delta V$ and the ratio of the young to the
old component $r_V$. The resulting variations in $r_V$
and $\Delta V$ are indicated in Figure 6 by shading and asterisks. It
is seen that the high-$r_V$ subdomains compose the spiral arm.

The distribution of high-$\Delta V$ subdomains
matches well with the visual impression of the $V$-band image
(Fig. 1a). The relation between $\Delta V$ and the
actual extinction $A_V$, however, is complicated, for
the absorbing matter is mixed with stars in the M31
disk. If we apply a simple model in which the absorption
layer is geometrically thin and lies just halfway along
the line of sight, $\Delta V\sim0.15$ may mean
complete opaqueness of the layer. Actual absorbing
clouds seem to be highly inhomogeneous, and the average
opaqueness derived from the shift of the luminosity
function may strongly reflect the optically thin parts.
In this sense, $\Delta V\sim0.15$ may also be
interpreted by a model in which a fraction 0.5 of the
field coverage is fully opaque and the other 0.5 is
completely transparent. Xu \& Helou (1996b) evaluated
the average extinction at $A_V\simeq1$ from the observed
far-infrared flux around $R=7$ kpc, integrated along the
line of sight. Therefore, when inhomogeneities are taken
into account, a significant fraction of the field of
view may effectively have a fully opaque layer.

We similarly introduce 15 subdomains in the
$3'\times3'$ field by combining
mesh fields according to the number of stars of $J=18 - 19$
in each mesh field. Each such subdomain has an area of about 0.6
arcmin$^2$, and the subdomains are numbered from 1 to 15;
see Figure 9. There again remain
mesh fields which are not combined into subdomains
because of relatively high local fluctuation. The $J$-
and $H$-band luminosity functions (Fig. 10) for the f/10
field show variations from one subdomain to another in
the ratio of the population for $H\sim17.6$ relative to
that for $H\sim18.2$,
$r_H\equiv\psi_H(17.6)/\psi_H(18.2)$. The subdomains in Figure 9 are
shaded according to the value of $r_H$. Judging from the
luminosity function for the $H$ band of the f/31 data,
which reaches down to $H=20.5$ (see Fig. 4b in Paper I),
these two magnitude ranges apparently reflect the
intermediate-age population ($t\sim1$ Gyr) and the older
ones ($t\sim10$ Gyr), respectively. Note, however, that this part of
the distribution functions of the f/10 data is already
subject to the selection effects due to the
observational limit. We additionally find a kind of
anticorrelation between the intermediate-age population
and the young population ($t<100$ Myr; $H<16.5$) among
the subdomains. The same characteristics are also found
in the $J$-band luminosity function, as shown in Figure
10a. It is most interesting to note that the subdomains
rich in the young population have less intermediate-age
population compared with the subdomains that are poor
in the young population. No clear sign of differential
extinction can be confirmed in the study of the $J$- or
$H$-band luminosity function for the subdomains. The
subdomains of different characteristics are indicated in
Figure 9 according to the value of $r_H$. The global trend of the
distribution of the subdomains rich in the young
population coincides between the $V$ study and the $JH$
study for the common $3'\times3'$ field.

\section{Discussion}

The present bolometric luminosity function (Fig. 3) well
resembles that in the M31 "disk" which was derived
by Rich et al. (1993) in their Field 2 at
$11.'5$ off center along the southeast minor
axis, except for the brightest end of $M_{bol}<-6$, which
reflects the additional youngest population in A24. The
slope is roughly $\Delta \log \psi/\Delta M_{bol}\sim1$
for $-3.5<M_{bol}<-6$ in both studies, although we have
noted the substructures in our case in Figure 3.

In applying the fuel consumption theorem of Renzini \&
Buzzoni (1986) for a single stellar population,
$n_j=B(t)L_Tt_j$, where $n_j$ is the number of stars in
post-main-sequence evolutionary phase $j$, $t_j$ is the
lifetime of the phase, $L_T$ is the total bolometric
luminosity of the population, and $B(t)=2\times 10^{-11}$
stars $L_\odot^{-1}$ yr$^{-1}$, we may examine the standard lifetime
estimate of $1.3$ Myr mag$^{-1}$ for AGB stars (Iben \& Renzini
1983). We use $\mu_K\sim16.5$ mag arcsec$^{-2}$ for the
M31 disk at $R\sim7$ kpc according to Battaner et al.
(1986) and $BC_K \sim 3.0$, to derive
\hbox{$L_T\sim10^6$ $L_\odot$} for the whole f/31
$2'\times2'$ field (effective area $\sim10,800$
arcsec$^2$). Although the population of the AGB stars of
$-5 \lesssim M_{bol} \lesssim -4$ are not of single age, we may apply
the above equation to the sum of the stars of intermediate
ages by virtue of its linearity in $L$ and $n$ when
$t_j$ mag$^{-1}$ is constant. For stars of $M_{bol}=-4.0$ and
$M_{bol}=-5.0$, we find $n_j\sim400$ and $n_j\sim50$,
therefore, estimates of $\log t_j\sim6.6$ and $\log t_j\sim 5.7$,
respectively. The simple average $\log t_j \sim6.15$ is
close to the standard value cited above ($\log t_j=6.1$),
basically conforming to the fuel consumption theorem.

When we apply the same fuel consumption theorem for
evaluating the lifetime of the luminous HROs
of $J-K \gtrsim 2$, which were identified in the f/31
$2'\times2'$ field by Kodaira et al. (1998b) as possible
candidates for the superwind-phase AGB stars, we find
$n_j\sim7$ in the f/31 $2'\times2'$ field and $\log
t_j\sim4.9$. When we take the uncertainty of the mixed
population noted above into account, this may be in
broad agreement with the estimate $\log t_j\sim5$ by
Tanabe et al.(1997) for similar objects in the
intermediate-age clusters in the LMC, to support their
inference about the high mass-loss rate of these
objects.

The present star formation rate in the observed
$7'\times7'$ domain over the last $\sim10$ Myr is
estimated at $\sim1.8\times 10^4$ $m_\odot$ Myr$^{-1}$ per
deprojected disk area of 1 kpc$^2$ for an inclination of
$i=12.^{\circ}5$. This estimate is derived from the number of
the young stars of $-4.5 \lesssim M_V \lesssim -3.5$ corresponding to
$\sim$B0 V stars of $\sim15 - 19$ $m_\odot$, $n=9$ arcmin$^{-2}$ mag$^{-1}$, by
adopting the Salpeter initial mass function for the mass
range of $0.1 - 120$ $m_\odot$.
As was pointed out in the last section, the ratio of the young population to
the old population varies among subdomains, suggesting local variation
in the star formation rate. The local intensification
factors of the star formation rate over the last $\sim10$
Myr are derived from the number of stars in the range
$-4.5 \lesssim M_V \lesssim -3.5$ and are estimated at
$\sim4$, $\sim2$, $\sim1$, and $\sim0.5$ times the above mean value
($\sim1.8\times 10^4$ $m_\odot$ Myr$^{-1}$ kpc$^{-2}$) in the subdomains
17, 15, 19, and 4, respectively (see Fig. 7).

Additionally, we obtain an estimate on the
average star formation rate $\sim7.2\cdot10^3$
$m_\odot$ Myr$^{-1}$ per deprojected area of 1 kpc$^2$ from
the observed star number $n=70$ arcmin$^{-2}$ mag$^{-1}$ of the
bright AGB stars of $-4.8 \lesssim M_{bol} \lesssim -3.8$ in the
$3'\times3'$ field. Thereby we assume that these stars have ages of
$\sim0.2 - 2.5$ Gyr (corresponding to the mass of
$\sim4.0 - 1.5$ $m_\odot$) and the standard lifetime of
$1.3$ Myr mag$^{-1}$. The bright AGB stars of $-4.8 \lesssim M_{bol} \lesssim
-3.8$ may have drifted from their birth volume because of the proper motion
and the resulting epicyclic motion, and their distribution is
correspondingly smeared out relative to the young stars with $-4.5
\lesssim M_V \lesssim -3.5$. The star formation rate derived from the
bright AGB stars may reflect the average rate in the disk at $R\simeq7$ kpc
rather than the specific domain related to A24. These star formation
rates slightly increase when the larger distance modulus ($m-M=24.47$)
or/and a stronger interstellar extinction would be
assumed. The above star formation rates for the M31 field around A24 may
be compared with the present star formation rates in the Galaxy,
$(3.5-5.0)\times 10^3$ $m_{\odot}$ Myr$^{-1}$ kpc$^{-2}$ for the 1 kpc vicinity of the
Sun, and $8\times 10^3$ $m_{\odot}$ Myr$^{-1}$ kpc$^{-2}$ for the disk in average at
$R \simeq 7$ kpc (see Rana 1991). This comparison suggests a similar
level of the present star formation activity between the average
Galactic disk and the M31 field around A24 on a spiral arm, and
consequently may indicate a slightly low star formation activity in the
M31 disk in average relative to the Galactic disk around $R=7$ kpc.
The relatively low present star formation activity in
M31 was pointed out by previous investigators (see
Tomita, Tomita, \& Sait\= o 1996 and references
therein). The present star formation rate of a whole galaxy was
estimated at $0.8-13$ $m_{\odot}$ yr$^{-1}$ for the Galaxy (Rana 1991 and
references therein) and at $0.2-0.5$ $m_{\odot}$ yr$^{-1}$ for M31 (Walterbos
1988), based upon various total emissions such as H$\alpha$ and the far-infrared.

A kind of anticorrelation between the young population
and the intermediate-age population, which was found in
the last section, may suggest that the star formation in
individual disk subdomains of a size $\sim100 - 200$
pc was not continuous but might have been intermittent,
like a "Christmas tree" model. An inspection of CMDs for the
different groups in Figures 10a and 10b suggests that the deficient AGBs
are mainly those of $t \lesssim 0.5$ Gyr. The history of the local star
formation may be smeared out over a timescale longer than this. The
rotation period of the disk domain of $R=4 - 10$ kpc is $\sim0.2$ Gyr,
and the period of the spiral-pattern passage is almost
the same in the case of a two-arm spiral (Braun 1991). Some
subdomains of the stellar complex A24 have just
experienced star formation in the present spiral passage
but did not in the last few passages. Some other subdomains
actively formed stars in the last passages one or two times
$\sim0.2$ Gyr ago but not in the present passage. Disk
domains of $\sim100 - 200$ pc size seem to be filled up
with cool gas sufficient to produce stars again on a
timescale of a few times the spiral passage period.

The actual dust and gas clouds in the studied field have
complex structures. The deep $V$-band image (Fig. 1a)
reveals fine complex textures of dark clouds surrounding
the clusters C202 and C203 in the superbubble, with
various scales of $1' - 0.'1$ ($200 - 20$ pc) down to
$1''$ ($\sim3$ pc) order of spotlike features that
might be cores or knots of large clouds. Individual
early-type stars show a wide range of interstellar
extinction of $A_V\simeq0 - 3$, again indicating highly
local inhomogeneities. The result of the analysis of the
local luminosity function in the $V$ band also supports
the inhomogeneous nature of the absorbing clouds. The
real three-dimensional structure of this stellar complex with
absorbing clouds is difficult to model uniquely
because of the awkward projection angle
$i=12.^{\circ}5$ of M31. The fine structures are not
particularly elongated along the major axis. Some
fibrous features are even oriented along the minor axis,
or perpendicular to the direction of the spiral arm.
Accordingly, we suspect that the dark cloud complex
surrounding the superbubble around C202 and C203 is not
perfectly flat in the central plane but is equally
extended above the plane, as was pointed out for dark
lanes in some edge-on spiral galaxies by Sofue (1987)
and Sofue, Wakamatsu, \& Malin (1994).

The spiral dark lane appears to be broken on both
sides of D221, but the ridge rich in young stars runs
through it along the spiral arm (Figs. 1 and 6).
Although the detailed three-dimensional structures of the clouds are
not clear, Figure 6 (see also Fig. 1b) gives an
impression for the global trend around the 7 kpc
spiral-arm structure that the distribution of the young
stellar population shows a wide ridge which seems to be
slightly shifted outward relative to that of the
absorbing clouds, conforming to the view about the
sequential distribution of the interstellar matter and
the newly born stars for the case of the trailing
spiral; this was suggested by Loinard et al. (1996) in
the study of the distribution of CO and HI clouds
relative to that of H II regions in the M31 spiral arms.

In conclusion, we believe that this work has demonstrated the possibility
of detailed population analysis of M31, and we would like to underline the
importance of extensive {\it deep} photometry in the visual as well as
in the infrared region.

\acknowledgments

The authors wish to acknowledge the collaborative
assistance of the staff members of the University of Hawaii
Institute for Astrophysics during the
observation at Mauna Kea, and the generous cooperation
in the $IRAS$ data processing by the Infrared Processing
and Analysis Center operated by CIT/JPL under the NASA
contract. We are grateful to A. Ku\v{c}inskas for
providing us with the data for deriving the bolometric
correction prior to publication. The computational
work was done using the facilities of the Astronomical
Data Analysis Computing Center of NAOJ. This work was
partially supported by Grant in Aid for Science Research
07640360.

\clearpage

\centerline{APPENDIX}

\appendix

\section{Bolometric Correction}

In order to transform $M_K$ to $M_{bol}$, we used the
bolometric correction $(BC_K)$ as a seventh-order polynomial
function of $(J-K)_0$:
$BC_K = \Sigma a_i(J-K)^i_0$, with the following coefficients:
$a_0=0.24$, $a_1=3.89$, $a_2=-1.795$, $a_3=0.586$, $a_4=-0.241$,
$a_5=0.0661$, $a_6=-0.00859$, $a_7=0.00041$, for
$-0.1 \leq (J-K)_0 \leq 3.5$. The polynomial was derived
by fitting to the empirical relation for giants in
globular clusters in the Milky Way [$-0.1 \leq (J-K)_0
\leq 1.4$] by Montegriffo et al. (1998), $BC_K$ values of
individual AGB stars in the South Galactic Cap [$1.1
\leq (J-K)_0 \leq 2.5$] from Whitelock et al. (1994,
1995), and $BC_K$ values of the OH/IR stars [$1.1 \leq (J-K)_0
\leq 3.5$] from Ku\v{c}inskas (1998), which were corrected
by $\Delta BC_K=+0.5$ to match others in the range
$1 < (J-K)_0 < 2$; see Figure 11. In the latter two
sources, the interstellar reddening $E(J-K)$ was
regarded as negligible compared with the intrinsic
dispersion. Although the sources were heterogeneous as to
the age and the metallicity, we believe that this
empirical correction incorporated the essential effects
of the extinction and the thermal emission by the hot
dust shells, and that this transformation is superior to
those used in previous works which assumed almost
constant $BC_K \simeq 3.25$ for $(J-K)_0 > 1.5$.

\section{Correction for the Dust-Shell}

The dust-shell structure and the radiative transfer
through the shell were calculated with the program DUSTY
by Ivezi\'{c}, Nenkova, \& Elitzur (1997). The density profile of the
stationary shell was calculated for the varying wind
velocity which was accelerated by the radiation pressure
on dust, and the dust temperature at the inner boundary
of the shell was taken to be 800 K. The dust opacity was
calculated for the $silicate$ parameters given in Draine
\& Lee (1984), in contrast to the special mixtures in
Bressan et al. (1998), and for the size distribution as
adopted by Mathis, Rumpl, \& Nordsieck (1977). Models
were calculated for the optical thickness of the shell
$\tau_V = 1$, 3, 10, 30, and 100, and for the central
blackbody source of temperature $T=2000$, $2500$, $3000$,
and $3500$ K. By comparing the original blackbody SED
with the SED of the corresponding shell model, we
evaluated the corrections in the near-infrared
photometric magnitudes. These corrections were applied
to the theoretical isochrones by the Padova group
(Bertelli et al. 1994) according to the following
conditions. The dust-shell was assumed to become
effective at $M_{bol} = -3.8$ with $\tau_V=1$. If this
point is too hot for relatively young populations (age
$100 - 200$ Myr), the first shell-effective point was
assumed to be the point on the original isochrone in the
$M_{bol}$ versus $(H-K)_0$ diagram (Fig. 5c), where the
isochrone deviates from the vertical path. The thickness
of the dust-shell was assumed to be $\tau_V=30$ at the
tip of the AGB, and the corrections were linearly
interpolated between the first point ($\tau_V=1$) and
the final one ($\tau_V=30$) proportional to $M_{bol}$.
The case of $\tau_V \leq 10$ for the tip of AGB turned
out to be insufficient, while the case of $\tau_V = 100$
produces too red AGB isochrones to match the
observation. The lower limit of $M_{bol} = -3.8$ and the
linear $M_{bol} - \tau_V$ relation were deduced from
the data of O-rich and C-rich stars (Ku\v{c}inskas 1998).

\clearpage

\centerline{FIGURE CAPTIONS}

Fig. 1.--- (a) $V$-band image showing the whole of the $7'\times 7'$ area
studied around A24. (b) Relative locations of the $7'\times 7'$
$V$ field, the $3'\times 3'$ $JH$ field, and the $2'\times 2'$ $JHK$ field.
The Hodge Atlas is underlaid. The frame of the $7' \times 7'$ field is slightly
inclined relative to others which are oriented with north
at the top and east to the left.

Fig. 2.--- $V$ luminosity function for the $7'\times 7'$
field. $A_V = 1.0$ is assumed.

Fig. 3.--- Bolometric luminosity function for the $2'\times 2'$ $JHK$ field.

Fig. 4.--- Two-color diagram for bright stars $(K<18.0)$ in
the $2'\times 2'$ $JHK$ field. The isochrones are those
modified using dust-shell models; see text. The ages
of the isochrones are 10, 20, 50, 100, 200, and 500 Myr
(AGB tracks with curved branching-off), and 5 and 10 Gyr
(AGB tracks with straight branching-off), but they are almost
degenerated. The oldest two cases are for
the helium and metal fractions $Y=0.25$ and $Z=0.008$,
and the others are for $Y=0.28$ and $Z=0.02$. Only the
RGB tracks for the oldest two cases extend to the
reddest part of $(H-K)_0>0.5$. A reddening vector for
$A_V = 3$ is indicated according to the Galactic
extinction law. Original isochrones of Padova 94 are
strongly degenerated, and that for the age of 100 Myr only
is shown in the inset with $IRAS$ AGB stars of Whitelock
et al. (1994, 1995).

Fig. 5.--- Color-magnitude diagrams for stars in the $2'\times 2'$ $JHK$ field.
The isochrones are those modified using dust-shell models; see text. The ages
of the isochrones are 10, 20, 50, 100, 200, and 500 Myr and
1, 2, 5, and 10 Gyr (from top to bottom). The oldest
two cases are for the helium and metal fractions
$Y=0.25$ and $Z=0.008$, and the others are for $Y=0.28$
and $Z=0.02$. A $dereddening$ vector for $A_V =5$ is
indicated in (a), according to the Galactic extinction
law. Original isochrones of Padova 94 are shown in the
insets for the same set of ages and metallicities as those in the main
frames.

Fig. 6.--- Subdomains in the $7'\times 7'$ $V$ field
for which luminosity functions were differentially
studied. The analyzed subdomains are numbered from 1 to 33 and are
indicated by a different number of asterisks and darkness according
to the extinction index $\Delta V$
and the population ratio index $r_V$: no asterisk means $\Delta V < 0.1$;
a single asterisk, $0.1 \leq \Delta V < 0.25$; two asterisks, $0.25 \leq
\Delta V \leq 0.5$. Decreasing degrees of darkness are for $r_V > 1.0$,
$0.5 \leq r_V \leq 1.0$, $0.25 < r_V \leq 0.5$, and $r_V \leq
0.25$, respectively. The clear mesh fields are not
combined into the subdomains. See text for the
definitions of $\Delta V$ and $r_V$.

Fig. 7.--- Typical examples of the magnitude distributions
for subdomains having different population ratio indices $r_V$:
subdomains 4 (circles), 19 (downward-pointing triangles), 15 (crosses),
and 17 (upward-pointing triangles) having $r_V=0.10$, 0.31, 0.58, and
1.1, respectively. The curves for subdomains 4, 19, and 15 are
vertically shifted relative to that of subdomain 17 by
multiplying by factors of 0.80, 1.09, and 1.03,
respectively. The straight lines show the gradients
0.5 and 2.24. See text for the definition of $r_V$.

Fig. 8.--- Typical magnitude distributions for subdomains
having different extinction indices $\Delta V$. See text for the
definition of $\Delta V$. (a) Subdomain $6$ (triangles) and
the reference subdomain 15 (circles). (b) Subdomain 26 (triangles)
and the reference subdomain 28 (circles). The curves for subdomains
6 and 26 are horizontally shifted by $\Delta V =-0.4$.

Fig. 9.--- Subdomains in the $3'\times 3'$ $H$ field
for which luminosity functions were differentially
studied. The analyzed subdomains are numbered from 1 to 15 and are
shaded according to the population ratio index $r_H$. Decreasing
degrees of darkness are for $r_H > 1$, $0.7 < r_H \leq 1$, and
$0.5 < r_H \leq 0.7$, respectively. The clear mesh
fields are not combined into subdomains. See text for
the definition of $r_H$.

Fig. 10.--- Typical magnitude distributions for subdomains
having different population ratio indices $r_H$. (a) $J$ and (b) $H$.
"High" subdomains $r_H \sim 1.0$ (Nos. 2, 4, 5, and 15; upward-pointing triangles),
"medium" subdomains $r_H \sim 0.7$ (Nos. 3, 10, 12, and 13;
circles), and "low" subdomains $r_H \sim 0.5$ (Nos. 7, 9, 11, and 14;
downward-pointing triangles). See text for the definition of $r_H$.

Fig. 11.--- Polynomial fitting of $BC_K$ as function of
$(J-K)_0$ to the empirical relation by Montegriffo et
al. (1998) (triangles), and the data of AGB stars in
Whitelock et al. (1994, 1995) (circles) and OH/IR stars
in Ku\v{c}inskas (1998) (dots). The Ku\v{c}inskas data were
shifted by 0.5 mag upward to match others in the
range $1<(J-K)_0<2$. In the latter two sources, the
reddening correction was assumed to be negligible
compared to the intrinsic dispersion.

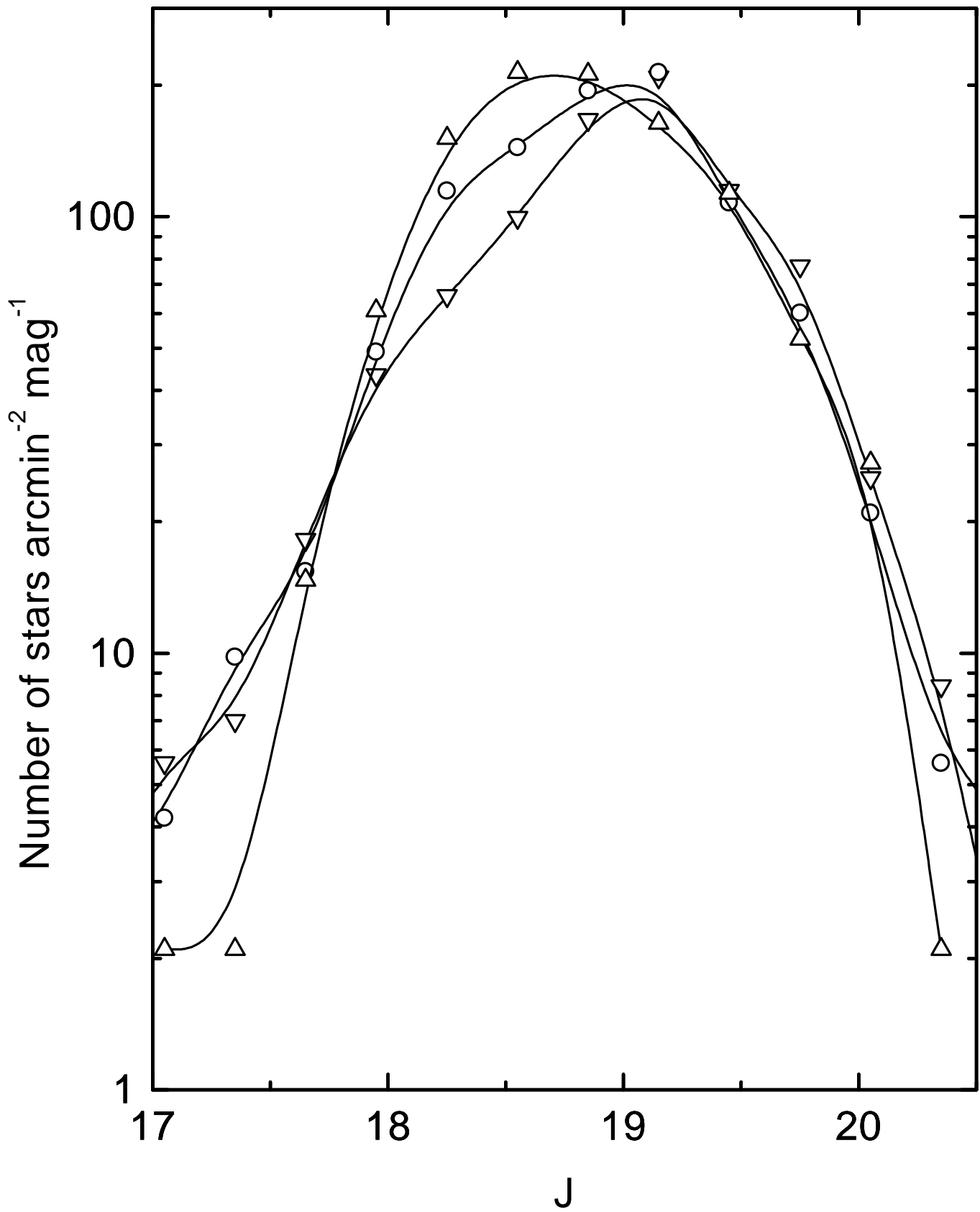
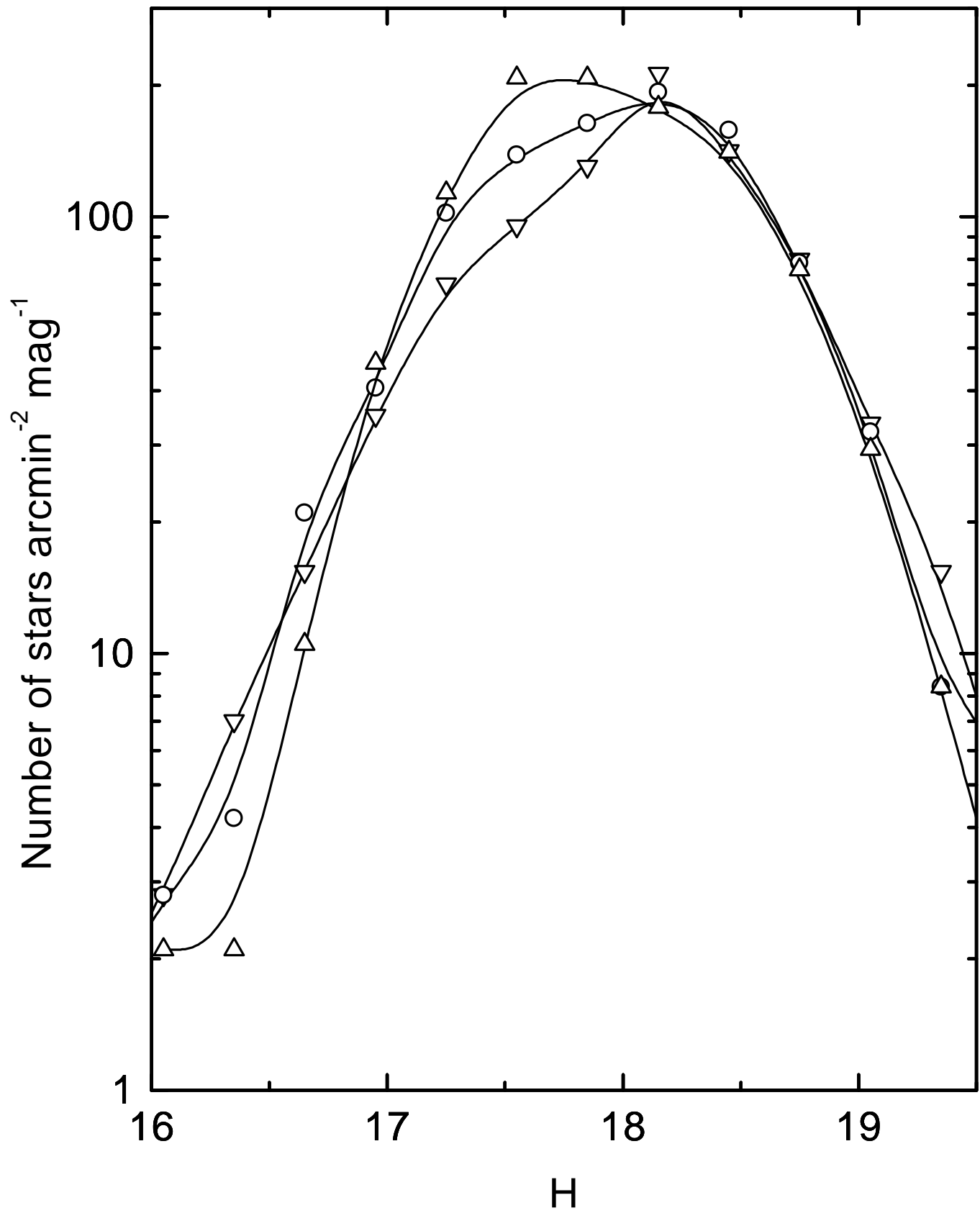
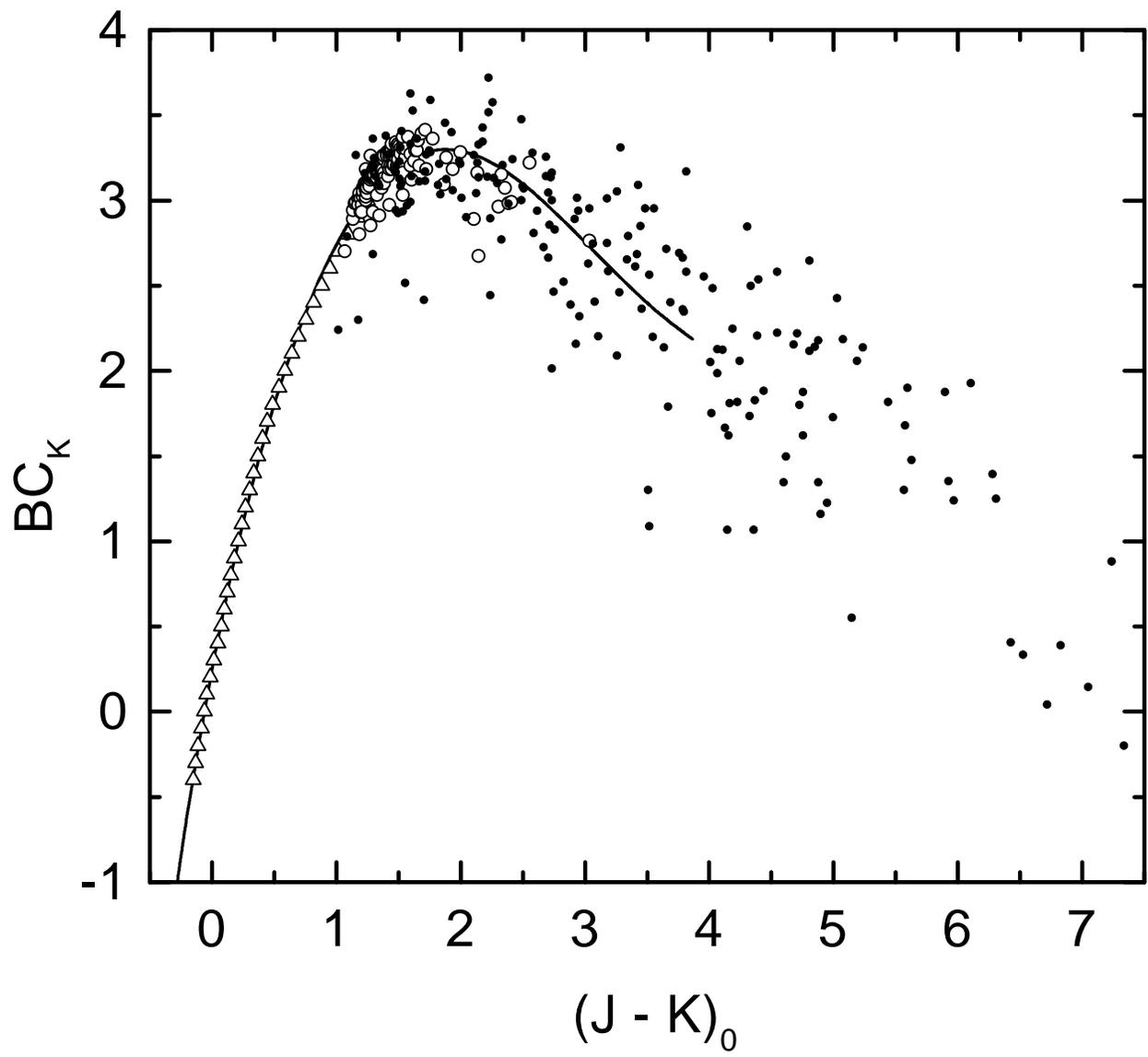
\end{document}